\documentclass[showpacs,aps,pra, preprint]{revtex4}

\usepackage{amsmath,graphicx,bbm,mathrsfs,amssymb,pst-all,bm,color}

\input{tcilatex}
\begin{document}

\title{Quantum algorithm for simulating the dynamics of an open quantum
system}
\author{Hefeng Wang, S. Ashhab and Franco Nori}
\affiliation{Advanced Science Institute, RIKEN, Wako-shi, Saitama 351-0198, Japan\\
Department of Physics, The University of Michigan, Ann Arbor, Michigan
48109-1040, USA}

\begin{abstract}
In the study of open quantum systems, one typically obtains the decoherence
dynamics by solving a master equation. The master equation is derived using
knowledge of some basic properties of the system, the environment and their
interaction: one basically needs to know the operators through which the
system couples to the environment and the spectral density of the
environment. For a large system, it could become prohibitively difficult to
even write down the appropriate master equation, let alone solve it on a
classical computer. In this paper, we present a quantum algorithm for
simulating the dynamics of an open quantum system. On a quantum computer,
the environment can be simulated using ancilla qubits with properly chosen
single-qubit frequencies and with properly designed coupling to the system
qubits. The parameters used in the simulation are easily derived from the
parameters of the system+environment Hamiltonian. The algorithm is designed
to simulate Markovian dynamics, but it can also be used to simulate
non-Markovian dynamics provided that this dynamics can be obtained by
embedding the system of interest into a larger system that obeys Markovian
dynamics. We estimate the resource requirements for the algorithm. In
particular, we show that for sufficiently slow decoherence a single ancilla
qubit could be sufficient to represent the entire environment, in principle.

\noindent
\end{abstract}

\maketitle

\section{Introduction}

For most problems of practical interest, the Schr\"{o}dinger
equation describing the evolution of a quantum system is too
complicated to be solved exactly. Numerical simulations are
therefore commonly employed for this purpose. However, simulating
quantum systems on a classical computer is a hard problem. The
dimension of the Hilbert space of the system increases
exponentially with the size of the system~(e.g., the number of
particles in the system). On a quantum computer, on the other
hand, the number of qubits required to simulate the system
increases linearly with the size of the system. As a result, the
simulation of quantum systems is more efficient on a quantum
computer~(see, e.g.,~\cite{bul, kas, somma1, somma2, somma3, aa,
whf1, temme}) than on a classical computer.

In general, quantum systems are never perfectly isolated from their
environments, and in some cases they must be treated as open systems.
Understanding the dynamics of open quantum systems is therefore important
for studying various quantum phenomena~\cite{gardiner, stolze, schl, weiss,
breuer1, cohen}. However, the simulation of open quantum systems on a
classical computer is also a hard task. In addition to the exponential
increase in the size of the Hilbert space of the open system, one also has
to consider the effect of the environment, which adds even more degrees of
freedom to the problem.

The Hamiltonian for an open quantum system coupled to an environment can be
expressed as
\begin{equation}
H=H_{S}+H_{B}+H_{I},
\end{equation}%
where $H_{S}$, $H_{B}$ and $H_{I}$ represent the Hamiltonians of the open
system, the environment and the interaction between the open system and the
environment, respectively. The interaction Hamiltonian $H_{I}$ can usually
be written in the form
\begin{equation}
H_{I}=\sum_{i}A_{i}\otimes B_{i},
\end{equation}%
where the operators $A_{i}$ and $B_{i}$ act in the state space of the open
system and the environment, respectively. In general, the environment has a
large number of degrees of freedom. Therefore the Hamiltonian in Eq.~($1$)
for the global system becomes too complicated to be solvable. However, one
is usually only interested in the evolution of the open system, and not the
environment. It turns out that for environments that have short correlation
times~(i.e., no memory effects) and induce slow decoherence in the system,
the microscopic details of the environment are not important. For purposes
of analyzing the dynamics of the system, one only needs to know a quantity
called the spectral density of the environment. The spectral density
characterizes the frequency distribution of the noise from the environment
on the open system. It combines the density of the environment modes and the
strength of the coupling between the environment modes and the system. For a
small set of discrete modes, the spectral density would consist of a
sequence of peaks. However, the frequencies of the environment modes are
usually so dense that the spectral density is a smooth function of frequency~%
\cite{weiss}.

Then, the task for the study of the dynamics of open systems can be
formulated as follows: Given the Hamiltonian of the open system, the
operators through which the open system interacts with the environment, the
spectral density of the environment and the temperature, how can we obtain
the dynamics of the open system? And how can we obtain the evolution of the
various physical properties of the open system?

In the case where the environment has no memory effects~(also referred to as
Markovian dynamics), the evolution of the system can be obtained by solving
the so-called master equation, which in the interaction picture has the form~%
\cite{breuer1}
\begin{widetext}
\begin{eqnarray}
\frac{d}{dt}\rho _{S}(t)\!&=&\!\sum_{\omega }\sum_{ij}\Biggl\{\!
-iS_{ij}\!\left(\omega \right)\!\left[A_{i}^{\dag }(\omega
)A_{j}(\omega),\rho_{S}\!(t)\right]\!+\!\gamma_{ij}(\omega
)\biggl\{A_{j}(\omega )\rho_{S}(t)A_{i}^{\dag }(\omega
)\!-\!\frac{1}{2}\left[A_{i}^{\dag }(\omega )A_{j}(\omega ), \rho
_{S}(t)\right]_{+}\biggr\}\Biggr\}.
\end{eqnarray}
\end{widetext} where $\rho _{S}(t)$ represents the density matrix of the
open system, and $A_{i}(\omega )$ is the decomposition of the operator $%
A_{i} $ into eigen-projectors of the Hamiltonian $H_{S}$. The operator $%
A_{i} $ represents the $i$-th operator that acts in the state space of the
open system in the interaction Hamiltonian as shown in Eq.~($2$). The
operators $A_{i}(\omega )$ can be very complicated in matrix form depending
on the details of the open system. The coefficients $S_{ij}\left( \omega
\right) $ are given~\cite{breuer1} by
\begin{widetext}
\begin{equation}
S_{ij}\left( \omega \right) =\frac{1}{2i}\left[\int_{0}^{\infty
}dte^{i\omega t}\langle B_{i}^{\dag }\left( t\right) B_{j}\left(
0\right) \rangle -\int_{0}^{\infty }dte^{-i\omega t}\langle
B_{i}^{\dag }\left( 0\right) B_{j}\left( t\right) \rangle \right],
\end{equation}
\end{widetext}where
\begin{equation}
B_{i}\left( t\right) =\exp \left( iH_{B}t\right) B_{i}\exp \left(
-iH_{B}t\right),
\end{equation}
and we have set $\hbar =1$. The operator $B_{i}$ represents the $i$-th
operator that acts in the state space of the environment in the interaction
Hamiltonian as shown in Eq.~($2$). The non-negative quantities $\gamma
_{ij}\left( \omega \right) $ play the role of decay rates for different
decay channels of the open system and are given in terms of certain
correlation functions of the environment~\cite{breuer1}
\begin{equation}
\gamma _{ij}\left( \omega \right) =\int_{-\infty }^{+\infty
}\!\!dt\;e^{i\omega t}\langle B_{i}^{\dag }\!\left( t\right) B_{j}\!\left(
0\right) \rangle .
\end{equation}

If one were to try and simulate the dynamics of a large open quantum system
on a classical computer, one would be faced with the problem that the number
of basis states grows exponentially with the size of the open system. The
master equation that describes the dynamics of the open system becomes too
complicated to be exactly solvable, and sometimes it is even practically
impossible to derive the master equation. One natural possibility for
tackling such problems is therefore to use quantum simulation.

There has been some work on the quantum simulation of open systems in the
literature. In Ref.~\cite{bacon}, the authors suggested an approach for
simulating the Markovian dynamics of an open quantum system on a quantum
computer. They showed that the simulation of the Markovian dynamics can be
reduced to building generators for a Markovian semigroup. However, as
mentioned above, the generator for the Markovian semigroup can be difficult
to obtain for a large system. In Ref.~\cite{terhal}, an approach for
preparing the thermal equilibrium state of an open quantum system was
suggested. Small-scale open system quantum simulators have also has been
demonstrated experimentally~\cite{bar, ryan}

In this paper, by extending the approach of Ref.~\cite{terhal}, we present a
quantum algorithm for simulating the Markovian dynamics of an open system
given the following information: the Hamiltonian of the open system, the
operators through which the open system interacts with the environment, the
spectral density of the environment and the temperature. This information
forms the input of the simulation.

The structure of this work is as follows: In Sec.~\ref{alg}, we present an
algorithm for simulating the dynamics of an open quantum system. In Sec.~\ref%
{res}, we estimate the resource requirements for the algorithm. In Sec.~\ref%
{eg}, we provide an example for the algorithm. We close with a conclusion
section.

\section{Theoretical description of the algorithm}

\label{alg}

In general, there are three steps in simulating the dynamics of an open
system on a quantum computer: first, preparing the initial state~(see, e.g.,~%
\cite{shende, berg, sok, nico, whf, bil}) of the open system and the
environment; second, implementing the dynamics on the open system and
finally reading out the state of the open system. We will focus on the step
of implementing the dynamics of the open system, and briefly describe the
other two steps of the algorithm in subsection II.C since they have been
discussed in the literature.

\subsection{Constructing a model Hamiltonian for the global system}

In general, the details related to the environment in Eq.~($1$)
are unknown. However, as mentioned above, one does not need to
know all these details in order to obtain the system dynamics. One
therefore has a good amount of freedom in constructing a model
Hamiltonian for the environment. One could therefore say that
under the Born-Markov approximation the spectral density is the
only piece of information that one needs to know about the
environment~(see e.g. Ref.~\cite{weiss, cohen}). Therefore, as
long as the spectral density of the model environment matches that
of the real environment, the effect on the system will be the
same.

In theoretical studies it is common to model the environment by a bath of
harmonic oscillators. For an open system in such an environment, the
environment Hamiltonian takes the form
\begin{equation}
H_{B}\!=\!\sum_{k}\omega _{k}\left( b_{k}^{\dag }b_{k}+\frac{1}{2}\right) ,
\end{equation}%
and the interaction Hamiltonian can be expressed as
\begin{equation}
H_{I}\!=\!\sum_{k}c_{k}\;\widetilde{A}_{k}\otimes \left(
b_{k}^{\dag }+b_{k}\right),
\end{equation}
where $b_{k}^{\dag }$~($b_{k}$) are the creation~(annihilation) operators of
the environment modes; $\omega _{k}$ are the frequencies of the environment
modes and $c_{k}$ are the coupling coefficients between the open system and
the environment modes; $\widetilde{A}_{k}$ are operators that act in the
state space of the open system and depend on the coupling mechanism between
the open system and the environment. For a set of discrete modes, the
spectral density is usually written as~\cite{weiss}
\begin{equation}
J\left( \omega \right) =\frac{\pi }{2}\sum_{k}\frac{c_{k}^{2}}{m_{k}\omega
_{k}}\;\delta \!\left( \omega -\omega _{k}\right) ,
\end{equation}%
where $m_{k}$ is the mass of the $k$-th oscillator. The $\delta $ in Eq.~($9$%
) is not restricted to infinitely-sharp $\delta $-functions, but to $\delta $%
-function approximants.

For purposes of simulating the dynamics of an open quantum system on a
digital quantum computer~(which is based on two-state qubits), it is
probably more natural to model the environment as a bath of two-level
systems or spin-$1/2$ particles. Such models are also sometimes used in
theoretical studies~(see, e.g.,~\cite{hanggi, lidar}). For an open system in
a spin bath, the environment Hamiltonian and the interaction Hamiltonian can
be expressed in the form
\begin{equation}
H_{B}=\frac{1}{2}\sum_{k}\omega _{k}\;\sigma _{k}^{z},
\end{equation}%
and
\begin{equation}
H_{I}=\frac{1}{2}\sum_{k}c_{k}\;\widetilde{A}_{k}\otimes \left(
g_{r}\,\sigma _{k}^{x}+g_{\varphi }\,\sigma _{k}^{z}\right) ,
\end{equation}%
respectively, where $\sigma _{k}^{x}$ and $\sigma _{k}^{z}$ are the Pauli
operators, $g_{r}$ and $g_{\varphi }$ are coefficients that describe the
relative size of the transverse coupling and the longitudinal coupling to
the open system. The transverse component induces relaxation, whereas the
longitudinal component induces pure dephasing. As will be explained in Sec.
II.D, there is a simple alternative method for simulating pure dephasing. We
therefore take $g_{r}=1$ and $g_{\varphi }=0$. The spectral density can then
be expressed as~\cite{hanggi}
\begin{equation}
J\left( \omega \right) =\pi \sum_{k}c_{k}^{2}\;\delta \!\left( \omega
-\omega _{k}\right) .
\end{equation}%
The difference between Eq.~($9$) and Eq.~($12$) is mostly a matter of
convention.

The environment mode frequencies $\omega _{k}$ and the coupling coefficients
$c_{k}$ can be determined as follows: We discretize the frequency spectrum
of the environment modes in the full frequency range from $\omega _{\min }$
to $\omega _{\max }$ into $d$ elements where each element has a width $%
\Delta \omega $:
\begin{equation}
\Delta \omega =\frac{\omega _{\max }-\omega _{\min }}{d}.
\end{equation}%
Correspondingly, the spectral density of the environment, $J\left( \omega
\right) $, is discretized and has the value $J\left( \omega _{k}\right) $
for the $k$-th element, where $\omega _{k}$ is the frequency of the $k$-th
element that represents the $k$-th environment mode. For a given $\omega
_{k} $, by making the following approximation
\begin{equation}
\int_{\omega _{k}-\Delta \omega /2}^{\omega _{k}+\Delta \omega /2}J\left(
\omega \right) d\omega \;\approx \;J\left( \omega _{k}\right) \Delta \omega
\;=\;\pi c_{k}^{2},
\end{equation}%
the corresponding coupling coefficient $c_{k}$ between the $k$-th element
and the open system can be obtained. Then the model Hamiltonian is
constructed based on the given information of the global system and will be
used in the implementation of the algorithm.

\subsection{Simulating the Markovian dynamics of an open system}

The dynamics of the open system is described by the evolution of the reduced
density matrix obtained by tracing out the environment degrees of freedom
from the density matrix of the global system:
\begin{equation}
\rho _{S}(t)=\text{Tr}_{B}[\rho (t)],
\end{equation}%
where $\rho _{S}(t)$ and $\rho (t)$ are the density matrices of the open
system and the global system, respectively. The density matrix $\rho (t)$
undergoes unitary evolution
\begin{equation}
\rho (t)=U(t,t_{0})\rho (t_{0})U^{\dag }(t,t_{0}),
\end{equation}%
where
\begin{equation}
U(t,t_{0})=\exp \left[ -iH(t-t_{0})\right] .
\end{equation}

Ideally, the dynamics of the open system can be obtained by coupling the
open system to an environment that has a large number of particles, letting
it evolve and reading out the state of the open system. In practice, on a
digital quantum computer that has a limited number of qubits, it is
impossible to represent all the particles of a typical environment.
Therefore we have to use an alternative technique to simulate the dynamics
of the open system in a large environment.

\begin{figure}[tbp]
\includegraphics[width=0.9\columnwidth, clip]{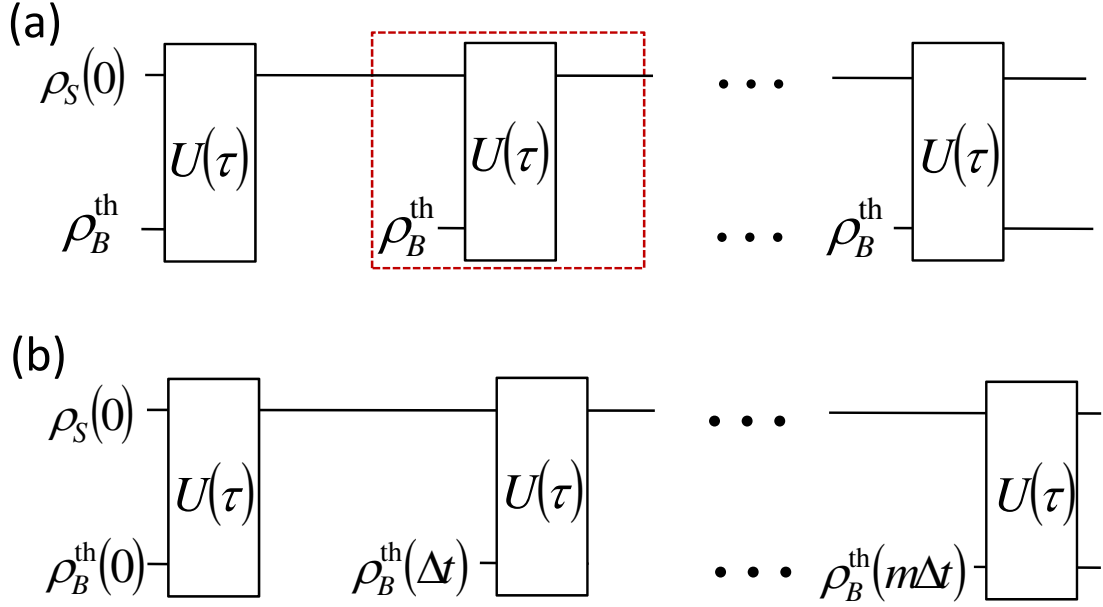}
\caption{(a)~Quantum circuit for simulating the Markovian dynamics of an
open quantum system coupled to a time-independent environment. The first
register represents the open system and the second register represents the
environment. (b)~Quantum circuit for simulating the Markovian dynamics of an
open system coupled to a time-dependent environment.}
\end{figure}

The large size of the environment plays a crucial role in justifying the
Markovian approximation. Under this approximation, the typical time during
which the internal correlations of the environment related to the effects of
the open system exist, $\tau _{R}$, is much shorter than the characteristic
relaxation time of the open system, $\tau _{S}$; as a result, the influence
of the open system on the environment is small and can be ignored.
Therefore, the state of the global system at any time $t$ can be
approximately described by the tensor product~\cite{breuer1}
\begin{equation}
\rho (t)\approx \rho _{S}(t)\otimes \rho _{B}^{\text{th}}.
\end{equation}%
where $\rho _{B}^{\text{th}}$ is the thermal equilibrium state of the
environment and can be written as
\begin{equation}
\rho _{B}^{\text{th}}=\sum_{j}P_{j}|j\rangle \langle j|,j=1,\cdots ,L=2^{d},
\end{equation}%
with
\begin{equation}
P_{j}=\frac{e^{-\beta E_{j}}}{Z},
\end{equation}%
and $|j\rangle $~($E_{j}$) are the eigenvectors~(eigenvalues) of the
environment Hamiltonian $H_{B}$, $Z$ is the partition function
\begin{equation}
Z=\sum_{j=1}^{L}e^{-\beta E_{j}},
\end{equation}%
where $\beta =1/k_{B}T$, $k_{B}$ is the Boltzmann constant and $T$ is the
temperature. Note that in the absence of interactions between the
environment qubits, the eigenstates $|j\rangle $ can be written as a tensor
product of the states of the environment qubits
\begin{equation}
|j\rangle =|j_{1}\rangle \otimes |j_{2}\rangle \otimes \cdots \otimes
|j_{d}\rangle ,
\end{equation}%
and the eigenvalues $E_{j}$ can be written as a sum of the eigenvalues of
the Hamiltonians of the environment qubits for the corresponding eigenstate $%
|j\rangle $%
\begin{equation}
E_{j}=E\left( j_{1}\right) +E\left( j_{2}\right) +\cdots +E\left(
j_{d}\right) .
\end{equation}%
As a result, the thermal-equilibrium state of the environment is a tensor
product of the thermal-equilibrium states of the individual qubits:%
\begin{equation}
\rho _{B}^{\text{th}}=\rho _{1}^{\text{th}}\otimes \rho _{2}^{\text{th}%
}\otimes \cdots \otimes \rho _{d}^{\text{th}},
\end{equation}%
where
\begin{equation}
\rho _{k}^{\text{th}}=\left( 1-p_{k}\right) |0\rangle \langle
0|+p_{k}|1\rangle \langle 1|,
\end{equation}%
denotes the thermal equilibrium state of the $k$-th environment qubit and%
\begin{equation}
p_{k}=\frac{1}{1+e^{\beta \omega _{k}}}.
\end{equation}

An alternative method that can be used to obtain the Markovian approximation
for a relatively small environment is to frequently force it back into its
thermal-equilibrium state. This process can be implemented relatively easily
on a quantum computer, where one has full access to all the qubits. The
procedure for simulating the Markovian dynamics of the open system is
therefore as follows: Set the environment to its thermal equilibrium state,
couple the open system to the environment modes and let the global system
evolve for some time, then reset the state of the environment to its thermal
equilibrium state. Repeat this process many times and then read out the
state of the open system. Mathematically, the evolution of the open system
in one step is expressed as
\begin{equation}
\rho _{S}^{\left( j+1\right) }\left[ (j+1)\tau \right] =\text{Tr}_{B}\left[
U(\tau )\rho ^{j}(j\tau )U^{\dag }(\tau )\right] ,j=0,1,2,\dots
\end{equation}%
where $\rho ^{j}(j\tau )=\rho _{S}^{j}\otimes \rho _{B}^{th}$.

Based on the above analysis, the evolution of the open system can be
simulated as follows on a quantum computer $\left[ \text{see Fig.}~1\left(
a\right) \right] $: prepare two quantum registers $R_{S}$ and $R_{B}$ to
represent the open system and the environment, respectively;

$\left( i\right) $ on the quantum register $R_{S}$, prepare the initial
state of the open system;

$\left( ii\right) $ on the quantum register $R_{B}$, prepare the thermal
equilibrium state of the environment;

$\left( iii\right) $ implement the unitary operation $U(\tau )=\exp (-iH\tau
)$ on the registers $R_{S}$ and $R_{B}$, where $H$ is the model Hamiltonian;

$\left( iv\right) $ repeat steps $\left( ii\right) -\left( iii\right) $ a
number of times.

$\left( v\right) $ read out the state, or the desired observable, of the
register $R_{S}$.

In general, the different terms in the model Hamiltonian do not commute. We
therefore employ the Trotter-Suzuki formula~\cite{nc} for implementing the
unitary operation $U(\tau )=\exp (-iH\tau )$
\begin{widetext}
\begin{eqnarray}
U(\tau ) &=&\exp \left[ -iH\tau \right]   \notag \\
&=&\lim_{n\rightarrow \infty }\left[ e^{-iH_{S}\tau /n}e^{-iH_{B_{1}}\tau
/n}e^{-iH_{I_{1}}\tau /n}\cdots e^{-iH_{B_{d}}\tau /n}e^{-iH_{I_{d}}\tau /n}%
\right] ^{n} \notag \\
&=&\lim_{n\rightarrow \infty }\left[ U_{S}\left( \tau /n\right)
U_{B_{1}}\left( \tau /n\right) U_{I_{1}}\left( \tau /n\right)
\cdots U_{B_{d}}\left( \tau /n\right) U_{I_{d}}\left( \tau
/n\right) \right] ^{n} \notag \\
&\approx&\left[ U_{S}\left( \tau /n_0\right) U_{B_{1}}\left( \tau
/n_0\right) U_{I_{1}}\left( \tau /n_0\right) \cdots
U_{B_{d}}\left( \tau /n_0\right) U_{I_{d}}\left( \tau /n_0\right)
\right] ^{n_0},
\end{eqnarray}
\end{widetext}where $n_{0}$ is a finite but large number.

In many cases, one is interested in the value of some physical properties of
the open system in the thermal-equilibrium state, such as various
correlation functions, the partition function, etc. The thermal equilibrium
state of the open system can be obtained by repeating the steps $\left(
ii\right) -\left( iii\right) $ until the open system reaches its thermal
equilibrium state~\cite{terhal}.

Note that the steps followed in implementing the algorithm explained above
are the same as those used in the algorithm of Ref.~\cite{terhal}. The goal
of that work, however, is different from the goal of our work. Our algorithm
simulates the dynamics whereas the algorithm of Ref.~\cite{terhal} aims to
prepare the thermal equilibrium state of the system. This difference in the
purpose of the algorithm leads to a number of further differences. For
example, in our case, we have to design the interaction Hamiltonian and the
parameters of the environment such that we accurately reproduce the spectral
density of the environment. In Ref.~\cite{terhal}, one only requires that
certain inequalities are satisfied in order for the environment register to
act as a good environment. In other words, the exact speed of reaching
thermal equilibrium is not a crucial issue in that work, as long as the
thermal-equilibrium state is reached in polynomial time. In contrast, if for
example there is some symmetry in the simulated system that prevents it from
reaching the thermal-equilibrium state, then our algorithm would still be
considered to work successfully if it produces the correct dynamics, even
though the thermal-equilibrium state is never reached.

\subsubsection{Timescales}

At this point we should make a few comments regarding the time scales
involved in applying the algorithm. The time interval $\tau $ should be very
short compared with $\tau _{S}$~(so that the state of the open system
changes slightly during $\tau $). The time $\tau $ can be considered the
memory time of the environment since the environment is reset to its thermal
equilibrium state at every time interval $\tau $. Since the timescale for
dynamics involving the system and a resonant mode of the environment is
given by the inverse of $c_{i}$ times a matrix element between energy
eigenstates of the system, the Markovian condition would require that $\tau $
must be small compared to this timescale, such that that the change in the
system's density matrix is small during the time $\tau $. When constructing
the model Hamiltonian, the frequency spectrum of the environment is
discretized into many elements where each element has a width $\Delta \omega
$. This width $\Delta \omega $ must be at most on the order of $1/\tau $ to
make sure that the different $\delta $-peaks have large overlap and produce
a smooth spectral density since the width of the $\delta $-peaks is on the
order of $1/\tau $.

Note that our algorithm can be straightforwardly generalized to simulate the
Markovian dynamics of an open system coupled to a time-dependent
environment~[see Fig.~$1\left( b\right) $]. In each iteration, the state of
the environment register is reset to the time-dependent thermal equilibrium
state of the time-dependent environment. The algorithm can also be
generalized for the case of an open system that is simultaneously coupled to
two or more different environments with different temperatures.

\subsubsection{Sequential application of the different dissipation channels}

In the above procedure for simulating the dynamics of an open quantum
system, we have assumed that each mode in the environment is represented by
one qubit. In the quantum circuit shown in Fig.~$1\left( a\right) $, the
environment quantum register $R_{B}$ is prepared in the thermal equilibrium
state of all the environment qubits with all the different frequencies. The
operator $U(\tau )$ acts in the state space of the system qubits and all the
environment qubits.

In this subsection, we show that one can reduce the number of qubits
required to represent the environment by having each qubit represent
multiple environment modes. In this approach each \textquotedblleft
evolve-reset\textquotedblright step in the algorithm above is split into
multiple evolve-reset steps, and in each one of those steps a qubit
represents one environment mode. However, the qubit is reset to different
frequencies in the different steps such that it produces the effect of
multiple environment modes on the system. Under certain conditions even a
single qubit can be used to represent the entire environment: the parameters
of this qubit are sequentially alternated between several different settings
such that the sequence covers all the dissipation channels of the
environment.

In the master equation shown in Eq.~($3$), the derivative of the reduced
density matrix of the open system is described by a sum over the decay
processes of the open system through all the different dissipation channels.
In the simulation algorithm, the decay of the open system in one step of the
evolution can be expressed as
\begin{eqnarray}
\rho _{S}^{\text{int}}\bigl[(j+1)\tau \bigr] &=&\text{Tr}_{B}\Bigl\{U^{\text{%
int}}(\tau )\bigl[\rho _{S}(j\tau )\otimes \rho _{B}^{th}\bigr]U^{\text{int}%
\dag }(\tau )\Bigr\}  \notag \\
&=&\left( 1-\tau \varGamma\right) \rho _{S}^{\text{int}}(j\tau ),
\end{eqnarray}%
where $\varGamma$ is a superoperator that describes the decay of the open
system and it is a sum over the contributions from all the different
environment modes~(the superscript \textquotedblleft int\textquotedblright\
indicates the interaction picture). Let $\varGamma_{j}$ denote the
superoperator that describes the decay of the open system caused by the
coupling of the open system to only the $j$-th environment mode. If the open
system undergoes a small change during the time $\tau $, such that $\tau %
\varGamma\ll 1$~(i.e., all the eigenvalues of $\varGamma$ times $\tau $ are
much smaller than $1$), then the evolution of the open system can be
approximated as
\begin{equation}
1-\tau \varGamma\approx \left( 1-\tau \varGamma_{1}\right) \left( 1-\tau %
\varGamma_{2}\right) \cdots \left( 1-\tau \varGamma_{d}\right) .
\end{equation}%
Therefore, the decay of the open system can be simulated by applying the
environment modes to the open system sequentially.

As a result, an alternative approach for simulating the dynamics of an open
system goes as follows: We divide the environment modes into a few sets and
let the different sets interact with the open system sequentially. In this
way, we effectively simulate the interaction between the open system and all
the different modes in the environment. The Hamiltonian for the open system
interacting with the $i$-th set of the environment modes is given by
\begin{equation}
H^{\left( i\right) }=H_{S}+\sum_{k=1}^{d_{i}}H_{B_{k}}^{\left( i\right)
}+\sum_{k=1}^{d_{i}}H_{I_{k}}^{\left( i\right) },
\end{equation}%
where $i=1,2,\cdots ,d/d_{i}$ and $d_{i}$ is the number of the environment
modes in each set.

One point that requires a little bit extra care here is that each mode in
the environment is coupled to the system in only one step out of $d/d_{i}$
steps, whereas the system Hamiltonian $H_{S}$ is applied in all of the
steps. One therefore needs to be careful how to calculate the elapsed time
in the simulated system. If the Hamiltonian in Eq.~($31$) is applied with a
given value of $\tau $, then after covering all the $d/d_{i}$ sets of
environment modes the system Hamiltonian would have induced a change
corresponding to a time $\tau \times d/d_{i}$, while each environment mode
would have induced a change corresponding to a time $\tau $. In order to
avoid any problems arising from this inconsistency, one can note that
decoherence rates are proportional to the spectral density~(and therefore
proportional to $c_{i}^{2}$) from the interaction Hamiltonian. One can then
use a rescaled Hamiltonian
\begin{equation}
\widetilde{H}^{^{\left( i\right)
}}=H_{S}+\sum_{j=1}^{d_{i}}H_{B_{j}}^{\left( i\right) }+\sqrt{\frac{d}{d_{i}}%
}\sum_{j=1}^{d_{i}}H_{I_{j}}^{\left( i\right) }.
\end{equation}%
If this Hamiltonian is now applied for a time $\tau $, then after covering
all the different environment modes both the system Hamiltonian and the
coupling to the environment would have induced changes that correspond to a
time $\tau \times d/d_{i}$, removing the inconsistency in the elapsed time.
Note that in order to guarantee the validity of the Markovian approximation,
the rescaled coupling strengths $c_{i}\times \sqrt{d/d_{i}}$ must still be
small compared to $1/\tau $.

The new procedure for simulating the Markovian dynamics of open systems is
implemented on a quantum computer as follows:

$i)$ on the quantum register $R_{S}$, prepare the initial state of the open
system;

$ii)$ on the quantum register $R_{B}$, prepare the thermal equilibrium state
of the qubits representing a subset of the environment modes;

$iii)$ implement the unitary operation $U_{i}(\tau )=\exp \left[ -i%
\widetilde{H}^{^{\left( i\right) }}\tau \right] $ on the registers $R_{S}$
and $R_{B}$;

$iv)$ perform steps $ii)-iii)$ for another set of environment modes, and
keep repeating this process until the algorithm runs over all the sets of
the environment modes;

$v)$ repeat steps $ii)-iv)$ a number of times;

$vi)$ read out the state, or the desired observable, of the register $R_{S}$;

In Eq.~($30$), the error in the decay of the open system introduced by
sequentially applying the dissipation channels, up to second order in $\tau $%
, is $\tau ^{2}\sum_{i,j}\varGamma_{i}\varGamma_{j}$. In the case of using
only a single qubit to represent the environment, the algorithm will have an
error of $d^{2}\tau ^{2}\overline{\varGamma}_{0}^{2}$, where $\overline{%
\varGamma}_{0}$ denotes the overall scale of the decay rate of the open
system caused by the coupling of the open system to a single environment
mode. When $d$ is large, this approximation can introduce a large error. On
the other hand, if we use more qubits to represent the environment, the
errors will be smaller because there will be fewer terms in the sum for the
error.

\subsection{State preparation and readout}

In the algorithm, the initial state of the global system is set to $\rho
_{S}(0)\otimes \rho _{B}^{\text{th}}$, where $\rho _{S}(0)$ is the initial
state of the open system and $\rho _{B}^{\text{th}}$ represents the thermal
equilibrium state of the environment. To prepare the initial state of the
global system on a quantum computer, we prepare two quantum registers $R_{S}$
and $R_{B}$ to represent the open system and the environment, respectively.
In general, the thermal equilibrium state of the environment is a mixed
state. In order to prepare the mixed state $\rho _{B}^{\text{th}}$ as shown
in Eq.~($19$), we can generate a random integer $j$, where $j\in \left[ 1%
\text{, }L\right] $, with probability $P_{j}$. Then we prepare the
corresponding state $|j\rangle $ on the quantum register $R_{B}$. We repeat
this step many times. This procedure produces an ensemble~(in time) of
states $|j\rangle $ with the corresponding probabilities $P_{j}$. This
ensemble gives the same effect as the case where the quantum register $R_{B}$
is prepared in a thermal equilibrium state in every time step.

As discussed in subsection II.A, in the absence of interactions between the
environment qubits, the thermal-equilibrium state of the environment is a
tensor product of the thermal-equilibrium states of the individual qubits as
shown in Eq.~($24$). In such cases, the thermal-equilibrium state of the
environment can be prepared in a simpler way: randomly generating $0$ or $1$
with respective probabilities $\left( 1-p_{k}\right) $ and $p_{k}$, then
preparing the corresponding states $|0\rangle $ or $|1\rangle $ on the
environment qubit, and repeating this step many times, the mixed state $\rho
_{B}^{\text{th}}$ can be prepared. In this procedure, an ensemble~(in time)
of states $|0\rangle $ and $|1\rangle $ with the corresponding probabilities
$p_{k}$ and $\left( 1-p_{k}\right) $ is produced, and it gives the same
effect as the case where the environment qubit is prepared in a thermal
equilibrium state in every time step.

\begin{figure}[tbp]
\includegraphics[width=0.9\columnwidth, clip]{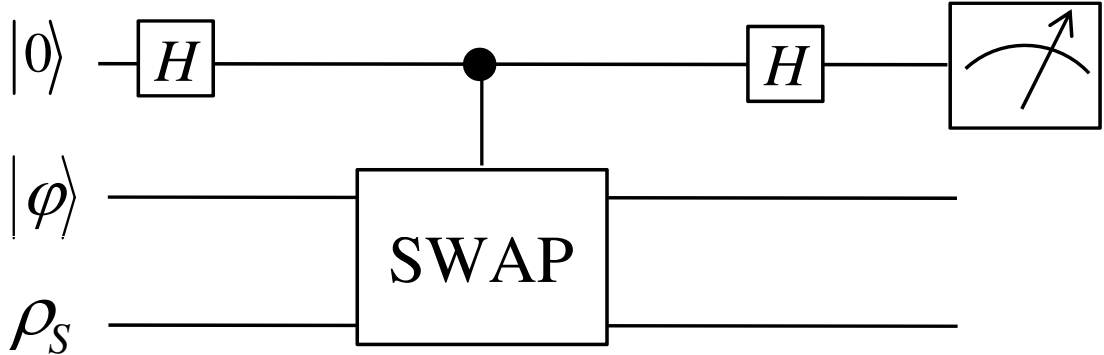}
\caption{Quantum circuit for a quantum estimator. $H$ represents the
Hadamard gate and SWAP represents the swap gate.}
\end{figure}

Observing the decoherence dynamics can be performed in a number of different
ways. For example, the phase estimation procedure~\cite{nc} can be used to
measure the energy of the open system, and one can then monitor the dynamics
of the energy distribution.

Alternatively, any matrix element in the density matrix of the open system
can be obtained using a quantum estimator~\cite{ekert, horodecki} as shown
in Fig. $2$. In the circuit for the quantum estimator, if we prepare the
second register in the state $|\varphi \rangle $ and the third register in
the state $|\psi \rangle $, then the value of $|\langle \varphi |\psi
\rangle |$ can be estimated by performing single-qubit measurements on the
index qubit. Through some derivation~\cite{ekert} we have
\begin{equation}
P\left( 0\right) =\frac{1}{2}\left( 1+|\langle \varphi |\psi \rangle
|^{2}\right) ,
\end{equation}%
where $P\left( 0\right) $ is the probability for obtaining the state $%
|0\rangle $ in the index qubit. If the third register is in a mixed state $%
\rho _{S}$, then we have
\begin{equation}
P\left( 0\right) =\frac{1}{2}\left( 1+\langle \varphi |\rho _{S}|\varphi
\rangle \right) .
\end{equation}

For the readout of the state of the open system in some chosen basis $%
\{|n\rangle \}$, any diagonal element $\rho _{nn}^{S}=\langle n|\rho
_{S}|n\rangle $ can be estimated by preparing the second register in the
state $|n\rangle $ and the third register in the state $\rho _{S}$. For the
off-diagonal elements $\rho _{mn}^{S}=\langle m|\rho _{S}|n\rangle $, the
real part of $\rho _{mn}^{S}$ can be estimated by preparing the second
register in the state $|\varphi \rangle =\left( |m\rangle +|n\rangle \right)
/\sqrt{2}$. The imaginary part of $\rho _{mn}^{S}$ can be estimated by
setting $|\varphi \rangle =\left( |m\rangle +i|n\rangle \right) /\sqrt{2}$.

The circuit shown in Fig.~$2$ can also be used for evaluating the
expectation values of arbitrary observables. For an operator $F$, by
applying the technique developed in Refs.~\cite{ekert, horodecki}, one can
obtain the value of $\langle F\rangle _{\rho _{S}}$. Then combined with the
quantum circuit for simulating the Markovian dynamics of open systems, one
can simulate the evolution of the expectation values of the physical
observables. Various correlation functions can also be obtained using this
technique.

\subsection{Simulating pure dephasing of a quantum system}

So far we have concentrated on the case where the nonunitary part of the
dynamics is caused by the environment modes that are resonant with the open
system. This picture is valid for energy relaxation. Pure dephasing, on the
other hand, is caused by low-frequency noise. Such low-frequency noise can
also be generated using the algorithm explained in the previous subsections:
whenever an environment qubit's state is changed from the ground to the
excited state or vice versa, the open system feels a telegraph-noise-like
change. This telegraph noise then causes~(mostly) pure dephasing in the
system. Although this effect can be induced using environment qubits, there
is a simpler method to generate telegraph noise. The telegraph noise is
essentially a classical noise signal affecting the open system. There is
therefore no need to use qubits in order to produce this classical signal.
It can be generated using a classical algorithm and added to the system
Hamiltonian $H_{S}$. If different noise signals are used in the different
runs of the algorithm, the density matrix of the system~(averaged over the
different realizations) will exhibit dephasing dynamics as a function of
time.

For an open system coupled to many fluctuators, the Hamiltonian for the
system can be expressed as~\cite{gal}~[see Eq.~($11$) for comparison]%
\begin{equation}
H=H_{S}+\sum_{k}\chi \!_{k}\left( t\right) \widetilde{A}_{k},
\end{equation}%
where
\begin{equation}
\ \chi _{k}\left( t\right) =\sum_{i}v_{ik}\;\xi _{ik}\left( t\right) ,
\end{equation}%
the random functions $\xi _{i}\left( t\right) $ characterize the
fluctuators' state, instantly switching between $\pm 1/2$ at random times.
Therefore, the simulation of the open system coupled to a bath of many
fluctuators is reduced to simulating a closed quantum system, which will
reduce the resources required for performing the simulation.

In simulating the dynamics of a quantum system coupled to many fluctuators,
we prepare the initial state of the system on a quantum register, then
implement the unitary operation $U=\exp \left( -iH\tau \right) $. Since the
signal $\chi \left( t\right) $ is random, to obtain the evolution of the
quantum system, we have to run the algorithm many times with a different
signal $\chi \left( t\right) $ in each run.

The noise spectrum of a noise signal is defined as~\cite{ymg}%
\begin{equation}
s\left( \omega \right) =\frac{1}{\pi }\int_{0}^{\infty }\!\!dt\;\cos \omega
t\langle \chi \left( t\right) \chi \left( 0\right) \rangle .
\end{equation}%
In order to generate the telegraph-noise signal~\cite{kog}, we
assume that the environment contains a number of fluctuators. Each
fluctuator switches between two possible configurations with
switching rate $\gamma _{i}$, and couples to the system with a
coupling strength $v_{i}$. The corresponding contribution of a
fluctuator to the noise spectrum is a Lorentzian~\cite{ymg},
\begin{equation}
s_{i}\left( \omega \right) =\frac{v_{i}^{2}\gamma _{i}}{4\pi \left( \omega
^{2}+\gamma _{i}^{2}\right) }.
\end{equation}
The~(low-frequency) noise spectrum of the entire environment is
the sum of the noise spectra of all the fluctuators
\begin{equation}
S\left( \omega \right) =\sum_{i}\frac{v_{i}^{2}\gamma _{i}}{\omega
^{2}+\gamma _{i}^{2}}.
\end{equation}
By adjusting the parameters $\gamma _{i}$ and $v_{i}$, one can produce a
variety of noise spectra. A typical example in practical situations is $1/f$
noise. If the number and density of fluctuators is sufficiently large, and
the distribution of the switching rates of the fluctuators $D\left( \gamma
\right) \propto \gamma ^{-1}$, and is independent of the distribution of the
coupling strengths between the fluctuators and the open system, the sum over
the fluctuators produces the $1/f$ noise spectrum~\cite{dutta}
\begin{equation}
S\left( \omega \right) =\frac{G}{\omega },
\end{equation}%
where $G$ is a constant.

\subsection{Simulating the non-Markovian dynamics of an open system}

The quantum algorithm presented above for simulating the Markovian dynamics
of an open system can also be used for simulating a class of non-Markovian
dynamics of open systems. A common situation where non-Markovian dynamics
occurs is the case where a small number of degrees of freedom in the
environment are coherent enough that they have non-negligible memory effects
in their interaction with the open system~\cite{breuer1}. Examples of this
situation include the relaxation dynamics of an atom through a cavity that
has a high quality factor~(see e.g., Ref.~\cite{breuer1}) and the relaxation
dynamics of a superconducting qubit close to resonance with a coherent
two-level defect~(see e.g. Ref.~\cite{ashhab}).

When a small number of degrees of freedom in the environment are responsible
for the non-Markovian dynamics, and assuming that one has sufficient
understanding of these degrees of freedom~(i.e. their intrinsic Hamiltonian,
their coupling to the system and their decoherence mechanisms and rates), it
becomes relatively straightforward to include them in the algorithm. One now
adds the appropriate number of ancilla qubits needed to describe these
degrees of freedom. When implementing the evolve-reset part of the
algorithm, one treats the additional degrees of freedom as part of the
system, i.e. they are not reset to their initial state. When the measurement
is performed at the end of the algorithm, however, only the system qubits
are used and the additional degrees of freedom are ignored. This step
corresponds to taking the trace over the state of these degrees of freedom.

The introduction of the additional degrees of freedom increases the resource
requirements~(which will be the subject of Sec.~III) as follows: the number
of qubits used in implementing the unitary operation $U(\tau )$ is increased
by $\log (D_{\text{ext}})$, where $D_{\text{ext}}$ is the number of degrees
of freedom in the environment that are responsible for the non-Markovian
dynamics. Since the interactions in the expanded system should still be
local, the scaling of resources will still be polynomial in system+ancilla
size and therefore efficient.

\subsection{Implementing environments other than spin baths}

Our algorithm simulates the effect of the environment using a set of ancilla
qubits that each represents one spin in a bath of independent spins~\cite%
{pro}. This type of environment is rather straightforward to implement,
since the properties and manipulation of the environment are done by
considering the ancilla qubits one at a time. For a variety of purposes,
this spin bath is sufficient for the implementation of the desired quantum
simulation. However, the spin bath has certain limitations. For example, the
two-level nature of the environment elements~(i.e., spins) means that
increasing the temperature will reduce the number of spins that are in their
ground states, thus reducing the relaxation rate induced by this environment~%
\cite{lidar}. This behavior contrasts with the case of a bath of harmonic
oscillators, where both relaxation and excitation rates increase with
increasing temperature. Therefore, if one is interested in the temperature
dependence of the dynamics, one needs to be careful about the differences
between different types of environments.

One possible technique to use a spin bath in order to simulate an oscillator
bath and obtain the correct temperature dependence is to calculate a
modified~(temperature-dependent) spectral density for the spin bath and use
this spectral density in the simulation. Alternatively, different types of
environments can be simulated by modifying the algorithm such that each
element in the environment is encoded into multiple qubits that are treated
as a single physical object. For example, one could use $n$ environment
qubits to represent the lowest $2^{n}$ energy levels of a harmonic
oscillator and then design a Hamiltonian where this harmonic oscillator
represents one mode in the environment. Thus, one can simulate the dynamics
of an open quantum system interacting with a bath of harmonic oscillators.
Note that the state preparation and the form of the Hamiltonian become more
complicated in this case than in the case of a spin bath.

\section{Resource estimation}

\label{res}

In this section, we discuss the resources including the number of qubits and
the operations needed for implementing the algorithm.

As shown in Fig.~$1\left( a\right) $, the number of qubits required for
simulating the Markovian dynamics of the open system is $\lceil \log
_{2}N\rceil +d$~(here $\lceil x\rceil $ provides the smallest integer larger
than $x$, or equal to $x$ if $x$ is an integer), where $N$ is the dimension
of the Hilbert space of the open system and $d$ is the total number of
qubits representing the environment. In the quantum estimator for the
readout of the state of the open system, $\lceil \log _{2}N\rceil +1$
additional qubits are needed as shown in Fig.~$2$. Therefore the total
number of qubits for simulating the Markovian dynamics of open systems is $%
2\lceil \log _{2}N\rceil +d+1$. One could also use the ancilla qubits that
are used in representing the environment in the readout of the state of the
open system, which would reduce the number of qubits used in the algorithm
to $\max \left[ \lceil \log _{2}N\rceil +d,2\lceil \log _{2}N\rceil +1\right]
$.

\begin{table}[tbp]
\caption{Resource needed for implementing the algorithm. Here $\lceil
x\rceil $ provides the smallest integer larger than $x$, or equal to $x$ if $%
x$ is an integer. $N$ is the dimension of the Hilbert space of the open
system, $d$ is the total number of qubits needed in representing the
environment in one time, $d_{i}$ is the number of qubits used in
representing the environment in approach $2$. $m$ is the number of times the
unitary operation $U(\protect\tau )=\exp (-iH\protect\tau )$ is implemented,
and $n_{0}$ is the parameter for dividing the time in the Trotter expansion
in Eq.~($28$).}
\begin{center}
\begin{tabular}{ccccc}
\hline
Algorithm &  & \# of qubits &  & \# of operations \\ \hline
Approach~$1$ &  & $2\lceil \log _{2}N\rceil \!+\!d\!+\!\!1$ &  & $O(m
\!\times \! \left(2d+1\right)^{n_{0}})$ \\ \hline
Approach~$2$ &  & $2\lceil \log _{2}N\rceil \!+\!d_{i}\!+\!\!1$ &  & $%
O(m\!\times\! d/d_{i}\!\times \!\left( 2d_{i}+1\right) ^{n_{0}})$ \\ \hline
\end{tabular}%
\end{center}
\end{table}

The unitary operation $U(\tau )=\exp (-iH\tau )$, where $H$ is the model
Hamiltonian, is implemented a finite number $m$ of times. To implement $%
U(\tau )$ on the quantum circuit, we employ the Trotter-Suzuki formula as
shown in Eq.~($28$), in which $U(\tau )$ is approximated by the product of $%
\left( 2d+1\right) ^{n_{0}}$ unitary transformations, where $n_{0}$ is the
parameter for dividing the time in the Trotter expansion. Therefore the
number of unitary operations needed in the quantum circuit shown in Fig.~$%
1\left( a\right) $ is $m\times \left( 2d+1\right) ^{n_{0}}$. Note one could
obtain higher efficiency using higher order Suzuki-Trotter formulas, as
discussed in Ref.~\cite{childs}

In the second approach we presented, the environment is represented using a
few or a single qubit. The number of qubits required for simulating the
dynamics of open systems is: $2\lceil \log _{2}N\rceil +d_{i}+1$, where $%
d_{i}$ is the number of qubits representing the environment.

The unitary operations implemented for each set of qubits are $U_{i}(\tau
)=\exp \left[ -i\widetilde{H}^{^{\left( i\right) }}\tau \right] $, where $%
\widetilde{H}^{^{\left( i\right) }}$ is given in Eq.~($32$). Thus
implementing $U_{i}(\tau )$ on the circuits by employing the Trotter-Suzuki
formula requires $\left( 2d_{i}+1\right) ^{n_{0}}$ unitary operations. For $%
d/d_{i}$ sets of the environment elements where each set of the elements is
implemented $m$ times, the total number of unitary operations that need to
be implemented is $m\times d/d_{i}\times \left( 2d_{i}+1\right) ^{n_{0}}$.

In preparing the initial state of the global system, we have to prepare the
thermal equilibrium state of the environment, which is a mixed state, a
number of times. To prepare the thermal equilibrium state $\rho _{B}^{th}$
as shown in Eq.~($19$), we prepare the state $|j\rangle $ with the
corresponding probability $P_{j}$, and repeat this step many times. In
running the algorithm, these basis states are fed to the register $R_{B}$ in
the quantum circuit in Fig.~$1\left( a\right) $ one at a time, and run the
algorithm many times with the the register $R_{B}$ prepared in the basis
states. For each basis state $|j\rangle $, we need to reset the state on the
register $R_{B}$ $m$ times in the algorithm. To do this, we need to erase
the state on the register $R_{B}$ and then prepare $R_{B}$ in state $%
|j\rangle $. We can first perform a measurement on the register $R_{B}$, the
state on $R_{B}$ collapse to a basis state, then we can perform a unitary
operation to rotate this state into the state $|j\rangle $.

In the readout of the state of the open system, we employ the quantum
estimator, in which the information of the open system is obtained by
performing single-qubit measurements on the index qubit and taking the
average. We have to prepare many copies of the state of the open system in
order to obtain accurate results. Both this procedure and the procedure for
preparing the thermal equilibrium state of the environment require running
the algorithm many times. The number of times for running the algorithm,
however, does not depend on the dimension of the Hilbert space of the open
system. Therefore the algorithm can be implemented efficiently using $O\left[
m\times \left( 2d+1\right) ^{n_{0}}\right] $ (or $O\left[ m\times
d/d_{i}\times \left( 2d_{i}+1\right) ^{n_{0}}\right] $, if the environment
is represented using $d_{i}$ qubits) unitary operations. All these results
are summarized in Table~I.

\section{Example: a two-level system in a spin bath}

\label{eg}

In this subsection, we consider the example of simulating the Markovian
dynamics of a two-level system that is immersed in a thermal bath of
independent two-level systems. The Hamiltonian of the global system is given
by%
\begin{equation}
H=-\frac{1}{2}\,\omega _{s}\,\sigma ^{z}-\frac{1}{2}\sum_{k}\omega
_{k}\,\sigma _{k}^{z}+\frac{1}{2}\sigma ^{x}\otimes \sum_{k}c_{k}\,\sigma
_{k}^{x},
\end{equation}%
where $\sigma ^{x}$ and $\sigma ^{z}$ are the Pauli operators, and $\omega
_{s}$ and $\omega _{k}$ are the frequencies of the two-level system and the
environment modes, respectively. The first term is the Hamiltonian of the
open system, the second term is the Hamiltonian of the environment and the
third term describes the interaction between the open system and the
environment.

In this example, assuming Ohmic dissipation, the spectral density of the
spin bath is expressed as~\cite{hanggi}
\begin{equation}
J\left( \omega \right) =2\pi \alpha \omega \exp \left( -\omega /\omega
_{c}\right) ,
\end{equation}%
where $\alpha $ is the dissipation coefficient and $\omega _{c}$ denotes the
cutoff frequency~($\omega _{c}/\omega _{s}\gg 1$). Below we specify
frequencies, temperatures and times using a standard unit frequency $\Delta
_{0}$. We first set $\omega _{s}/\Delta _{0}=1$, $\alpha =2\times 10^{-4}$, $%
\omega _{c}/\Delta _{0}=100$, and $\beta \Delta _{0}=1$. The frequency
spectrum in the region $\omega /\Delta _{0}\in \left[ 0.8\text{, }1.15\right]
$ is discretized into eight elements at frequencies $\omega _{k}/\Delta
_{0}=\left( 0.80+0.05k\right) $, with $k=0$, $1$, $\cdots 7$. The width of
each element is $\Delta \omega /\Delta _{0}=0.05$. The coupling coefficients
$c_{k}$ are determined using Eq.~($14$).

For this example, the analytical results for the relaxation rate and the
dephasing rate, $1/T_{1}$ and $1/T_{2}$, can be derived under the Markovian
approximation~\cite{cohen} as
\begin{equation}
\frac{1}{T_{1}}=\frac{1}{2}J\left( \omega =\omega _{s}\right) ,\text{ \ \ \
\ }\frac{1}{T_{2}}=\frac{1}{2T_{1}}.
\end{equation}%
The Markovian dynamics of the two-level system is simulated with the initial
state of the open system being set to the excited state $|e\rangle $ of the
two-level system. We set the unitary evolution time $\Delta _{0}\tau =30$,
and we obtain the evolution of the state of the open system. The relaxation
dynamics of the two-level system is shown in Fig.~$3$.
\begin{figure}[tbp]
\includegraphics[width=0.9\columnwidth, clip]{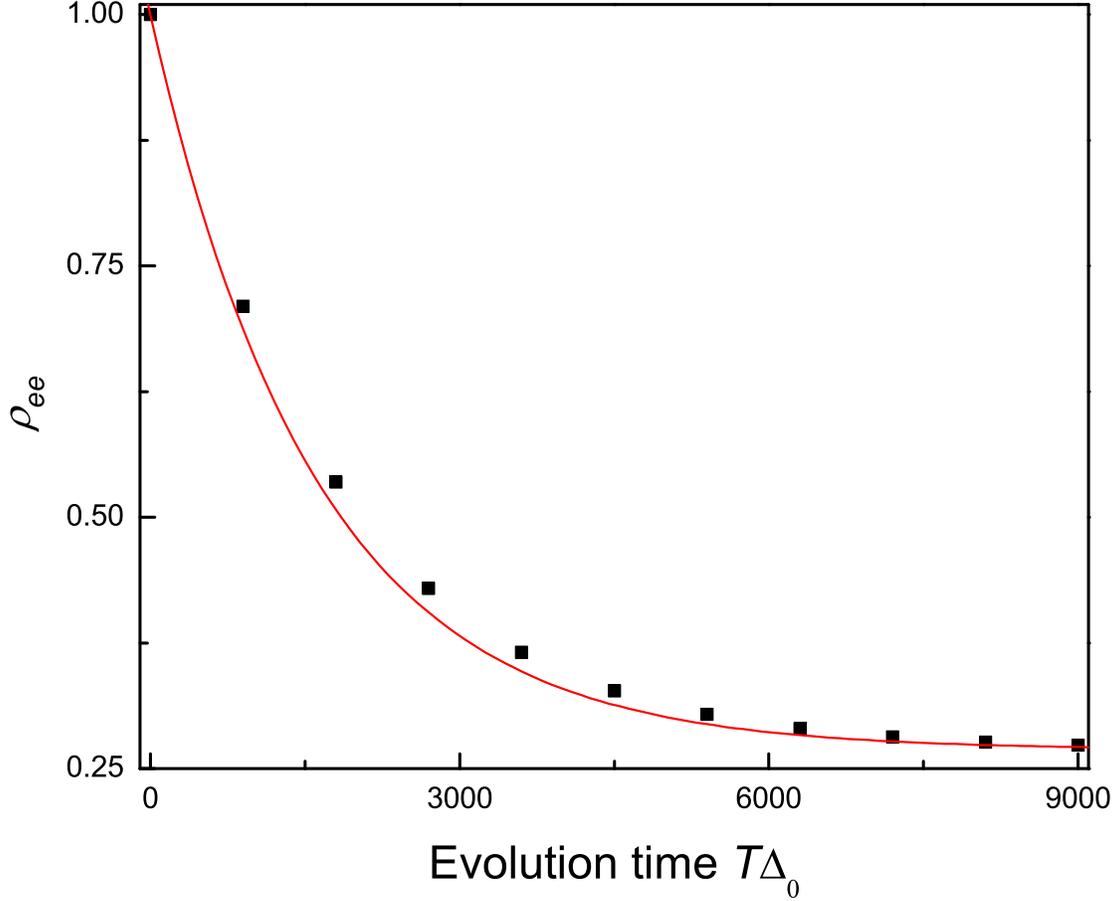}
\caption{(Color online)~The evolution of the matrix element $\protect\rho %
_{ee}$ of a two-level system in a spin bath, where $\protect\rho _{ee}$
denotes the diagonal element of the density matrix that describes the
population of the excited state. The frequency of the environment mode in
the region $\protect\omega /\Delta _{0}\in \left[ 0.8\text{, }1.15\right] $
is divided into $8$ elements with equal width and each element is
represented by a qubit. The Hamiltonian for the global system is given by
Eq.~($41$). The unitary evolution time $\Delta _{0}\protect\tau =30$. The
red solid line represents the analytical result for the evolution of $%
\protect\rho _{ee}$ with the initial state $|e\rangle $. The black square
dots represent the simulated results for the evolution of $\protect\rho %
_{ee} $.}
\end{figure}
\begin{figure}[tbp]
\includegraphics[width=0.9\columnwidth, clip]{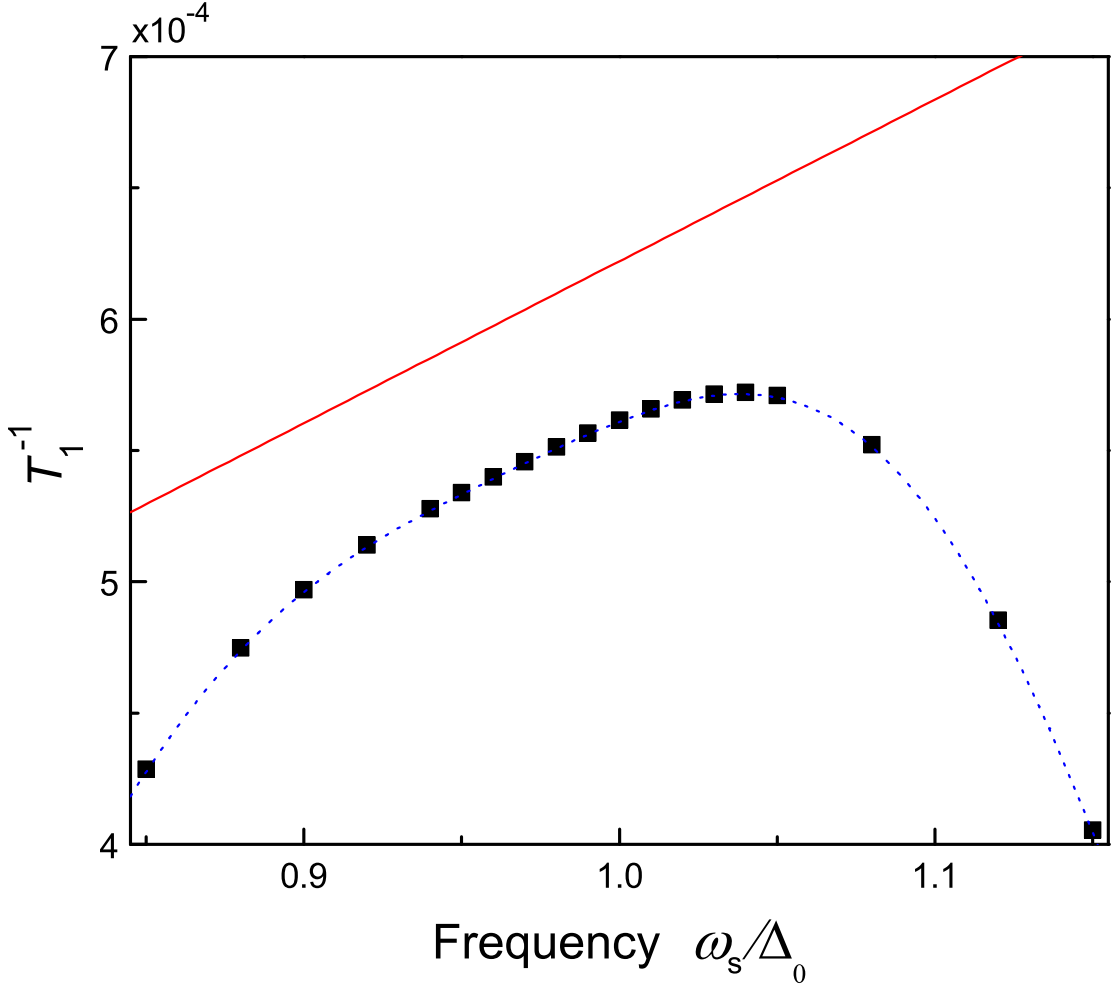}
\caption{(Color online)~The relaxation rate of the two-level system as a
function of its frequency. The black squares represent the results obtained
from the simulation of the algorithm; the blue dotted line represents the
relaxation rate obtained by using the spectral density that is spanned by
eight $\protect\delta $-peaks in Eq.~($44$); and the red solid line
represents the relaxation rate obtained from the analytical spectral density
in Eq.~($42$). Obviously only the first two ones agree well with each other.}
\end{figure}
\begin{figure}[tbp]
\includegraphics[width=0.9\columnwidth, clip]{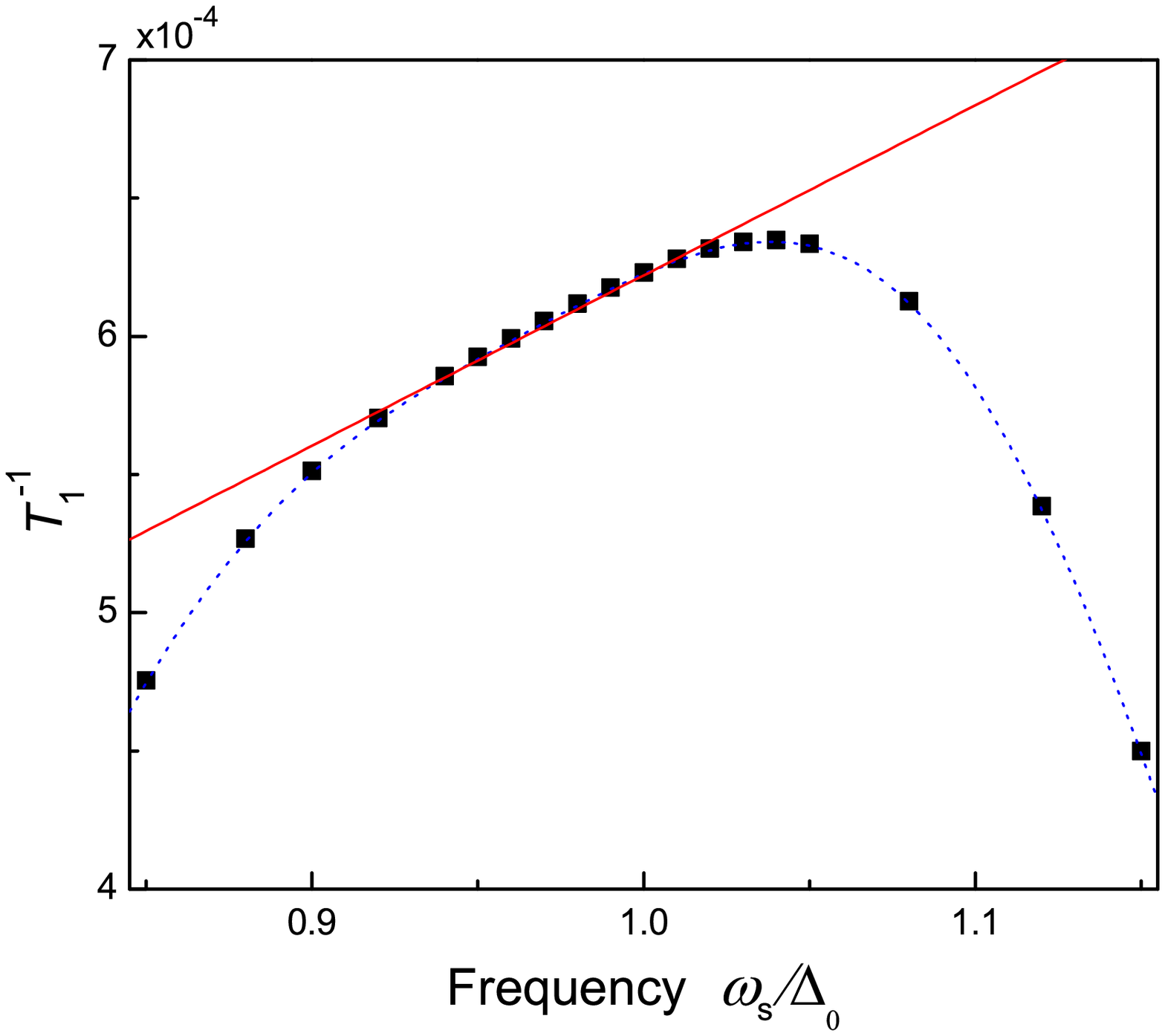}
\caption{(Color online)~The relaxation rate of the two-level system as a
function of the frequency of the two-level system. The black square dots
represent the results obtained from the simulation of the algorithm using
the improved coupling coefficients as discussed in the text; the blue dotted
line represents the relaxation rate obtained by using the spectral density
that is produced by eight $\protect\delta $-peaks in Eq.~($44$) using the
improved coupling coefficients; and the red solid line represents the
relaxation rate obtained from the analytical spectral density in Eq.~($42$).}
\end{figure}

From Fig.~$3$, we can see that there is a clear discrepancy between the
relaxation rates obtained from the numerical simulation and the exact
results given by Eq.~($43$). This discrepancy is due to the fact that we
used Eq.~($14$) to determine the coupling coefficients $c_{k}$ even though
only eight $\delta $-peaks are used to represent the spectral density:
\begin{equation}
J\left( \omega \right) =\pi \sum_{k=1}^{k=8}c_{k}^{2}\;\delta \left( \omega
-\omega _{k}\right) .
\end{equation}%
Each $\delta $-peak has the analytical form~\cite{terhal}%
\begin{equation}
\delta \left( \frac{\omega -\omega _{0}}{\Delta _{0}},\tau \Delta
_{0}\right) =\frac{1-\cos [\tau \left( \omega -\omega _{0}\right) ]}{\pi
\tau \Delta _{0}\left[ (\omega -\omega _{0})/\Delta _{0}\right] ^{2}}.
\end{equation}%
In Eq.~($14$), a good approximation can be achieved when many $\delta $%
-peaks are used to represent the spectral density and the peaks cover a
sufficiently wide range of frequencies. The eight peaks used in our
simulation do not satisfy these conditions.

In order to further demonstrate the above explanation of the discrepancy, in
Fig.~$4$ we show the relaxation rate as a function of the frequency $\omega
_{s}$ of the two-level system. There we compare the numerical results
obtained from the simulation, the analytical results obtained from Eq.~($43$%
) with the spectral density given by Eqs.~($44$) and ($45$) and the coupling
coefficients determined through Eq.~($14$); and the exact results obtained
from Eq.~($43$) with the spectral density given by Eq.~($42$). From Fig.~$4$
we can see that the numerical results for the relaxation rate obtained from
the simulation fit very well the analytical results of the eight-peak
spectral density, while both have a systematic deviation from the exact
results.
\begin{table}[tbp]
\caption{Results for simulating the Markovian dynamics of a two level system
using different number of qubits to represent the environment to
sequentially simulate a total of eight dissipation channels. $N$ denotes the
number of qubits representing the environment modes. The exact result for $%
T_{1}$ is $T_{1}^{\text{exact}}=2/J(\protect\omega =\Delta _{0})$, and $%
T_{2}^{\text{exact}}=2T_{1}$. The results here are obtained using the
improved choice of the coupling coefficients as discussed in the text. The
results are essentially independent of $N$.}
\begin{center}
\begin{tabular}{ccccccccc}
\hline
$N$ &  & $1$ &  & $2$ &  & $4$ &  & $8$ \\ \hline
$T_{1}/T_{1}^{\text{exact}}$ &  & $0.998$ &  & $0.998$ &  & $0.998$ &  & $%
0.996$ \\ \hline
$T_{2}/T_{1}^{\text{exact}}$ &  & $1.994$ &  & $1.990$ &  & $1.991$ &  & $%
1.991$ \\ \hline
\end{tabular}%
\end{center}
\end{table}

This systematic deviation from the exact results can be eliminated by
plugging in the analytical form of the $\delta $-peaks as shown in Eq.~($45$%
) into Eq.~($44$) for the eight elements and then obtaining an improved
approximation for the coupling coefficients $c_{k}$. In Fig.~$5$, we show
the relaxation rate as a function of the frequency of the two-level system
using the improved choice of the coupling coefficients. One can see that in
the region around $\omega _{s}/\Delta _{0}=1.0$, the numerical results are
now in good agreement with the exact results.

We also perform the same simulation with the sequential application of the
different dissipation channels. In different simulations we use different
numbers of qubits to represent the environment. We use $1,2,4,$ or $8$
qubits to represent the environment using the improved choice of $c_{k}$.
The ratios $T_{1}/T_{1}^{\text{exact}}$ and $T_{2}/T_{1}^{\text{exact}}$ are
shown in Table~II. One can see that they are in good agreement with the
exact results and that the different simulations give essentially the same
results.

We do not perform any numerical calculations for the simulation of pure
dephasing using telegraph noise here, because such calculations would follow
closely similar calculations that have been performed in the literature in
theoretical studies of telegraph-noise-induced dephasing~(see, e.g., Ref.~%
\cite{ymg, fal}).

\section{Conclusion}

\label{con}

In this paper, we have presented an algorithm for simulating the Markovian
dynamics of an open quantum system. The algorithm takes as an input the
Hamiltonian of the open system, the operators through which the open system
interacts with the environment, the spectral density of the environment and
temperature. One therefore does not explicitly deal with the master equation
describing the dynamics. In the simulation, the environment is represented
by a set of ancilla qubits that are designed to have the same effect on the
open system as the simulated environment. We have also shown that different
dissipation channels can be implemented sequentially, thus reducing the
number of qubits needed to represent the environment. Pure dephasing also
allows a reduction in the number of needed qubits, since it can be induced
by a properly designed classical noise signal. The algorithm can also be
used to simulate non-Markovian dynamics.

In the present algorithm, the ancilla qubits play a rather passive role in
the sense that they only facilitate the dissipative dynamics of the system.
These ancilla qubits could be used in a more active role as probes or
actuators for the open quantum system. By monitoring the response of these
ancilla qubits as they interact with the open system, one could obtain the
energy spectrum of the system. Once the spectrum is known, the ancilla
qubits can also be used to provide or absorb any given amount of energy and
guide the system to any desired energy eigenstate. The details of this
algorithm will be presented elsewhere.

\begin{acknowledgements}
We thank L.-A. Wu for helpful discussions. We acknowledge partial
support from DARPA, Air Force Office for Scientific Research, the
Laboratory of Physical Sciences, National Security Agency, Army
Research Office, National Science Foundation grant No.~0726909,
JSPS-RFBR contract No.~09-02-92114, Grant-in-Aid for Scientific
Research~(S), MEXT Kakenhi on Quantum Cybernetics, and Funding
Program for Innovative R\&D on S\&T~(FIRST).
\end{acknowledgements}

\end{document}